\documentclass[aps,showpacs,preprint,superscriptaddress]{revtex4}
\usepackage{graphicx}
\usepackage{amsmath}
\usepackage{array,color}
\usepackage{bm}

\begin{document}

\title{QED cascade induced by high energy $\gamma$ photon in strong laser field }
\author{Suo Tang}
\author{Muhammad Ali Bake}
\affiliation{Key Laboratory of Beam Technology and Materials Modification of the Ministry of Education, and College of Nuclear Science and Technology, Beijing Normal University, Beijing 100875, China}
\author{Hong-Yu Wang}
\affiliation{Department of Physics, Anshan Normal University, Anshan 114005, China}
\author{Bai-Song Xie\footnote{Corresponding author. Email address: bsxie@bnu.edu.cn}}
\affiliation{Key Laboratory of Beam Technology and Materials Modification of the Ministry of Education, College of Nuclear Science and Technology, Beijing Normal University, Beijing 100875, China}

\begin{abstract}
The QED cascade induced by the two counter-propagating lasers is studied. It is demonstrated that the probability of a seed-photon to create a pair is much larger than that of a seed-electron. By analyzing the dynamic characteristics of the electron and positron created by the seed-photon, it is found that the created particles are more probable to emit photons than the seed-electron. With these result, further more, we also demonstrate that the QED cascade can be easier to be triggered by the seed-photon than by the seed-electron with the same incident energy and the same laser.
\end{abstract}

\maketitle

\section{Introduction}
\label{secI}

Because of the latest technical and experimental progress, the multipetawatt and exawatt optical laser systems are being conceived \cite{RMP}. In the extreme light infrastructure (ELI) \cite{Eli} and the high power laser energy research (HiPER) facility \cite{HiPER}, the ultrahigh intensities exceeding $10^{25}\mathrm{W/cm^{2}}$ are envisaged. At such intensity, many phenomena in quantum electrodynamics (QED) \cite{W.Greiner} can be realized in laboratory such as photon splitting \cite{Adler} and electron-positron pair creation \cite{tang}. And recently, several works predicted that the ultra-relativistic laser could induce the QED cascade which would be an abundant source of electrons, positrons and photons.

The pair creation, which is one of the most fascinating phenomena in strong field QED, had been demonstrated by many experimental works \cite{PRL1626,PRL105001,PRL105003} through several different processes, for example, Bethe-Heitler process \cite{Heitler}($\gamma+Z\rightarrow e^{-}+e^{+}+Z$) which is important only in the materials with high-$Z$ nucleus, Breit-Wheeler process \cite{Breit} ($\gamma+n\omega_{laser}\rightarrow e^{-}+e^{+}$) which dominates in the ultrahigh intensity laser, the trident process \cite{Heitler,080401} ($e^{-}+n\omega\rightarrow 2e^{-}+e^{+}$) in which a pair is created in the interaction of the real photons with a virtual photon from the field of the charged particle.

If the QED cascade is realized in laboratory, it would open a new way to several fields in physics including antimatter science \cite{456}, pair plasma physics \cite{2333,085014} and astrophysical phenomena \cite{2092}. In this paper, we study the QED cascade in the strong laser field with the practical envelop function Eq.\ (\ref{eq7}). The cascade can be described as a positive feedback as follows. The electrons and positrons are first accelerated by the laser and emit hard photons through the non-linear Compton scattering ($e^{-/+}+n\omega_{laser}\rightarrow e^{-/+}+\gamma$) \cite{PRL76}, then, the emitted photons interact with the laser field to create pairs through the Breit-Wheeler process, and the new created electrons and positrons are accelerated in the laser field and new photons are produced again for the new creation process. In these events the laser not only plays the role as a target for the pairs and photons, but also acts as the accelerator for the charged particles. In our calculation, we do not take the trident process into account because the trident process is a second order process proportional to $\alpha^{2}$ \cite{Heitler}, where $\alpha$ is the fine structure constant.

The QED cascade triggered by seed-electron or positron in the standing wave formed by the counter-propagating lasers with the same intensity and polarization has been studied \cite{085008,080402}, but, in practice, it is not easy for the charged particle to move into the inner of the ultra-relativistic laser because the charged particle would be expelled by the extremely strong ponderomotive force, which will decrease the probability for emission. To overcome the ponderomotive force, the incoming particles need to be accelerated to ultra-relativistic energy region \cite{1582,195005}, for example, the electron beam with incident energy $46.6 GeV$ is applied in the famous experiment at SLAC \cite{PRL76,PRL1626,PRD092004}. The other approach to avoid the ponderomotive force is to replace the charged seed by a high-energy photon, which would not be influenced by the electromagnetic field. And the new electron and positron created by the seed-photon would be within the laser field naturally.

The high-energy photon and electron are applied as the seeds to trigger the QED cascade in our work. We found that the probability of a seed-photon to create a pair is much larger than that of a seed-electron, and the electron and positron created by the seed-photon are more likely to emit photons than the seed-electron. Thus, to some extent, these results can be applied to estimate the probability to trigger a cascade by the seeds and we demonstrate that the cascade can be easier to be triggered by the seed-photon than by the seed-electron.

We use a Monte Carlo method to simulate the quantum process \cite{015009} including Breit-Wheeler process and nonlinear Compton scattering, which is described in Sec.\ref{secII}. In Sec.\ref{secIII}, we compare the probability of the different kind of seeds to create a pair and in Sec.\ref{secIV}, the probability of the particles created by the seed-photon and of the seed-electron is studied by analyzing the moving parameters of them. In Sec.\ref{secV}, we study the cascades triggered by different seeds and compare the increasing rates of the particles created in the cascade. Finally, in Sec.\ref{secVI}, we provide a summary of our work.

\section{theoretical approach}
\label{secII}

Three dimensionless parameters $a$, $\eta$, $\chi$ determine the characteristics of the QED cascade. As is well known, $a=eE/mc\omega$, which is the parameter indicating the intensity of the laser, can be interpreted as the work performed by the laser field on the electron (positron) in one laser wavelength in the units of electron mass in our situation \cite{RMP}. The value of $\eta$ measures the importance of the nonlinear quantum effect and indicates the amplitude of the field in the rest frame of electron (positron) in the units of critical field $E_{crit}$ \cite{J.Schwinger}. If $\eta >1$, the nonlinear Compton scattering and quantum recoil are important in the motion of the charged particles. The Breit-Wheeler process is considerable only if $\chi >1$, which can be considered as the measurement of the field in the center-of-mass system of the created electron and positron \cite{RMP}.

Concretely the parameter $\eta$ and $\chi$ can be expressed as
\begin{align}
\eta&=\frac{e \hbar}{m^{3}c^{4}}\mid F^{\mu\nu} P_{\nu} \mid\notag \\
    &=\frac{\gamma}{E_{crit}}\sqrt{(\bm{E}+\bm{v}\times \bm{B})^{2}-(\bm{E}\bm{v}/c)^{2}},
\label{eq1}
\end{align}
and
\begin{align}
\chi&=\frac{e \hbar^{2}}{m^{3}c^{4}}\mid F^{\mu\nu} K_{\nu} \mid\notag \\
    &=\frac{\varepsilon}{E_{crit}}\sqrt{(\bm{E}+c\hat{k}\times\bm{B})^{2}-(\bm{E}\hat{k})^{2}},
\label{eq2}
\end{align}
where $e$ and $m$ is the charge and mass of the electron (positron), $c$ is the speed of light, $\gamma$ and $\varepsilon$ is the energy of electron (positron) and photon in the units of electron mass, $P_{\nu}$, $K_{\nu}$ is the four-momentum of electron (positron) and photon, $\bm{v}$ is the velocity of the electron (positron) and $\hat{k}$ is the unit vector of the photon, $F^{\mu\nu}$ is the electromagnetic field tensor and $\bm{E}$ and $\bm{B}$ are the electric and magnetic field of the laser.

The probability rate for emitting a photon with energy $\varepsilon mc^{2}$ by a electron (positron) with energy $\gamma mc^{2}$ can be written as \cite{Landau,035001}
\begin{align}
\emph{d}W_{em}(\xi)&=\frac{\alpha mc^{2}}{\sqrt{3}\pi\hbar\gamma}[(1-\xi+\frac{1}{1-\xi})K_{2/3}(\delta)\notag \\
                   &-\int_{\delta}^{\infty}K_{2/3}(s)\emph{d}s]\emph{d}\xi,
\label{eq3}
\end{align}
where $\xi=\varepsilon/\gamma$, $\delta=2\xi/[3(1-\xi)\eta]$ and $K_{2/3}(s)$ is the modified Bessel function of fractional order. The probability rate for creating a pair by a photon with energy $\varepsilon mc^{2}$ can be expressed as \cite{Landau,035001}
\begin{align}
\emph{d}W_{pair}(\xi_{-})&=\frac{\alpha mc^{2}}{\sqrt{3}\pi\hbar\varepsilon}[(\frac{\xi_{-}}{\xi_{+}}+\frac{\xi_{+}}{\xi_{-}})K_{2/3}(\delta)\notag \\
&-\int_{\delta}^{\infty}K_{2/3}(s)\emph{d}s]\emph{d}\xi_{-},
\label{eq4}
\end{align}
where $\xi_{-}=\gamma/\varepsilon$ and $\xi_{+}=1-\xi_{-}$ are the energy of the created electron and positron in units of the parental photon energy, and $\delta=2/3\chi\xi_{+}\xi_{-}$.

In principle, the probability rates above in the electromagnetic field also depend on two Lorentz invariant parameters $f=( E^{2}-c^{2}B^{2})/E^{2}_{crit}$ and $g=c(\bm{E}\cdot \bm{B})/E^{2}_{crit}$, but in our situation the laser field, which will be shown below, can be approximated as the plane wave, thus the dependence of the parametera $f$ and $g$ can be neglected as $f \approx 0$ and $g\approx 0$. On the other hand the probability rates above are calculated for a constant electromagnetic field, which is not a real laser field for our studied problem, however, if the variant time $T_{laser}$ of the laser field is much longer than the coherence time $t_{ch}$ of the quantum effects for the studied problem, i.e., $T_{laser} \gg t_{ch}=mc/eE$, the probability rates in the constant field can be regarded as a good approximation in our studied situation. Obviously the condition is satisfied naturally in the ultra-relativistic laser field since $a \gg 1$.

In our calculation we trace the motion of each electron, positron and photon. The electron (positron) obeys the classical equation of motion
\begin{align}
\frac{\emph{d}\bm{p}}{\emph{d}t}&=q(\bm{E}+\bm{v}\times \bm{B})\nonumber\\
\frac{\emph{d}\bm{r}}{\emph{d}t}&=\bm{v}
\label{eq5}
\end{align}
in each calculational time step. Obviously the motion of the photon is not influenced by the electromagnetic field but it can decay. After each step, we calculate the probability for emission, i.e., $W_{em}\emph{d}t$. If $r_{1}<W_{em}\emph{d}t$, where $r_{1}$ is a uniformly random number between $0$ and $1$, a photon will be emitted by the electron (positron). And the energy of the emitted photon is obtained by solving the equation below
\begin{equation}
r_{2}=\frac{\int^{\xi^{f}}_{0}\emph{d}W_{em}}{\int^{1}_{0}\emph{d}W_{em}},
\label{eq6}
\end{equation}
where $r_{2}$, ($0<r_{2}<1$), is acquired randomly. Then, we can obtain the energy of the photon as $\varepsilon^{f}=\xi^{f}\gamma$, where $\gamma$ is the energy of the parental electron(positron). The direction of the motion of the photon is parallel to the momentum $\bm{p}$ of the parental electron (positron), thus the final momentum of the parental electron (positron) can be written as $\bm{p}^{f}=\bm{p}-\varepsilon^{f}mc \bm{p}/p$. In fact the photon may be emitted over a cone of opening angle $1/\gamma$ approximately \cite{015009}, which is not taken into account in our paper.

The calculating process for the decay of photon is similar as above except the decayed photon would not be calculated at next step as the electron (positron) does. It is worth to note that the time step $\emph{d}t$ must be chosen such that $W\emph{d}t\ll1$, otherwise the result would depend on the value of the time step.

\section{the probability of the different seeds to create a pair}
\label{secIII}

In order to simulate the realistic situation in experiment, we employ two counter-propagating monochromatic laser with circular polarization and model the electric and magnetic fields of the pulses with envelope function $f_{\pm}(\phi_{\pm})$ for the rightward and leftward waves \cite{085008} as that
\begin{equation}
f_{\pm}(\phi_{\pm})=\frac{1}{4}[1\mp \tanh(\frac{\phi_{\pm}}{W_{la}})][1\pm \tanh(\frac{\phi_{\pm}\pm L_{la}}{W_{la}})],
\label{eq7}
\end{equation}
where $L_{la}=20$ is the length of the laser field, $W_{la}=2$ is the thickness of the pulse edges and $\phi_{\pm}=z\mp t$. By the way, here and throughout this paper we have used the phase units, i.e., $\omega t\rightarrow t$, $kx, ky, kz \rightarrow x, y, z$. For the transverse profile of the laser pulse we multiply the fields by the Gaussian envelope $\exp(-(x^{2}+y^{2})/\sigma^{2})$, where $\sigma=\pi$ is the half of the wavelength. Hence the field would be much weaker if the position is far away from the axis. The pulse of wavelength $\lambda=1\mathrm{\mu m}$ is chosen, thus the duration of the pulse is about $11\mathrm{fs}$. The field strength of the pulse with peak intensity $I$ is $E=\sqrt{I/c\epsilon_{0}}$ and $B=\sqrt{\mu_{0}I/c}$, where $\epsilon_{0}$ and $\mu_{0}$ are the permittivity and permeability of vacuum, respectively. The calculation starts at $t_{0}=-20$ with a time step $\emph{d}t=0.0001$ and ends at $t=40$, thus the initial and final distance between the two pulses is about $40$.

\begin{figure}[htbp]\suppressfloats
  \includegraphics[width=8cm]{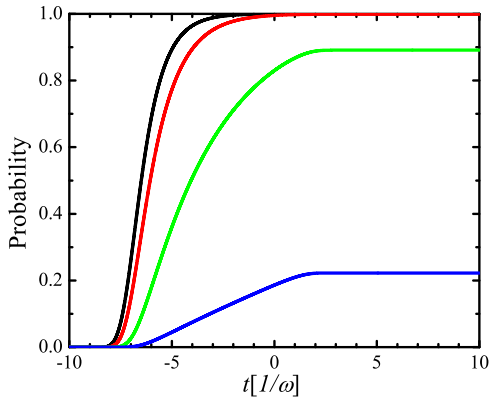}
  \caption{(Color online) The probability of a seed-photon to create a pair with $I=5\times10^{24}\mathrm{W/cm^{2}}$. The incident energy is $\varepsilon=600$ (black line),$400$ (red line), $200$ (green line), $100$ (blue line). At $t=-7.5$ the photon encounters the leftward laser, then the photon has the probability to create a pair. At $t=2.5$ the photon passes through the laser.}\label{fig.1.}
\end{figure}

At each time step, the probability of the seed-photon for creation is $W_{pair}\bm{d}t$, thus we can obtain the creating probability after $n$ time steps as $p=1-\prod^{n}_{i=1}(1-W^{i}_{pair}\bm{d}t)$ which is shown in Fig.\ref{fig.1.}. The photon propagates to the right from the initial position $z=-5$. As we can see, at $t=-7.5$ the photon encounters the leftward laser, then the photon has the probability to create a pair, and the probability stops increasing after the photon travels through the laser at $t=2.5$. For the photon with incident energy $\varepsilon=600$ propagating in the laser $I=5\times10^{24}\mathrm{W/cm^{2}}$, it must create a pair because the probability has reached $1$ before passing through the laser. However, the photon with incident energy $\varepsilon=100$ can hardly create a pair during the interaction.

\begin{figure}[htbp]\suppressfloats
  \includegraphics[width=6cm]{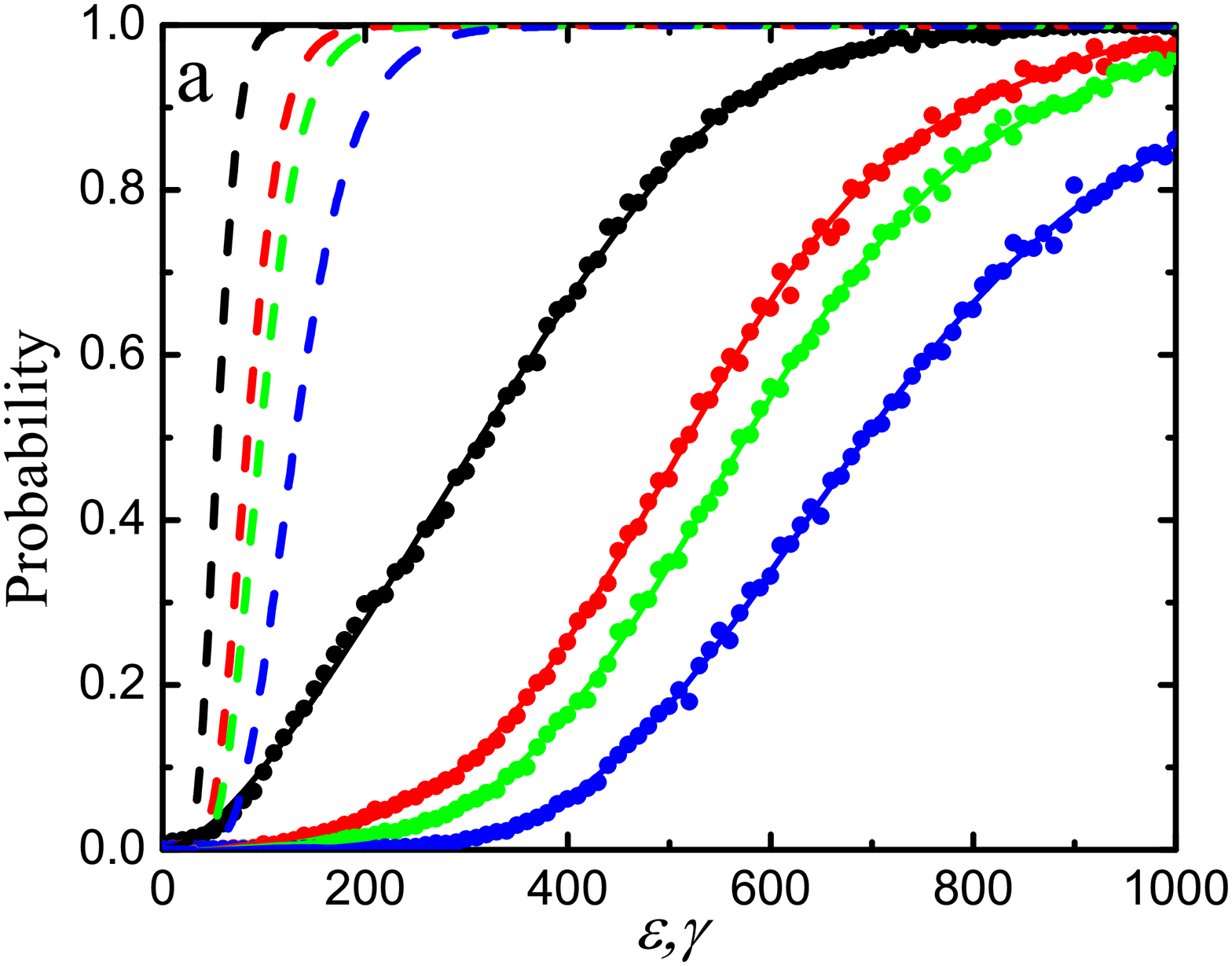}
  \includegraphics[width=6cm]{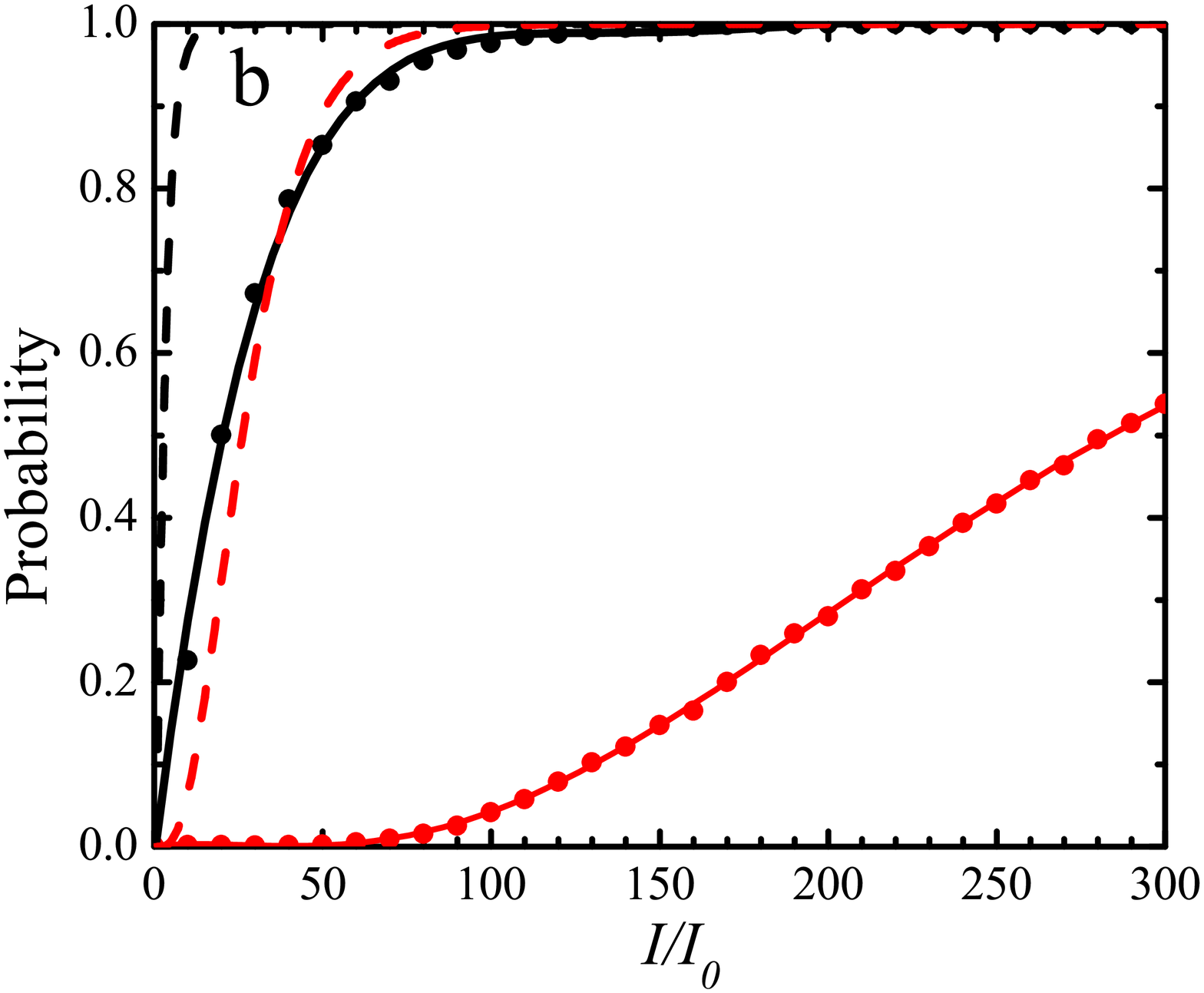}
  \caption{(Color online) The probability to create a pair by different seeds with different incident energy and different laser intensity. The dashed lines show the probability of a seed-photon. The dot line indicate the statistic probability of electrons for a large number ($10^{5}$) and the solid lines are the polynomial fitting. In (a), the seeds interact with $I=2\times10^{25}\mathrm{W/cm^{2}}$ (black line), $I=1\times10^{25}\mathrm{W/cm^{2}}$ (red line), $I=8\times10^{24}\mathrm{W/cm^{2}}$ (green line), $I=5\times10^{24}\mathrm{W/cm^{2}}$ (blue line). In (b), the incident energy of the seeds is $1000$ (black line), $200$ (red line), and the intensity is scaled to match $I_{0}=10^{23}\mathrm{W/cm^{2}}$.}\label{fig.2.}
\end{figure}

In Fig.\ref{fig.2.}, we show the final probability for pair creation, see the dashed lines, of the seed-photon with different incident energy which interacts with different intensity of laser. By the way, the rightward pulse does not interact with the photon so that the probability only depends on the interaction between the photon and the leftward pulse, but we still include the right pulse in the calculation in order to compare with the result of the seed-electron, which is shown by the dots in Fig.\ref{fig.2.}. The calculation process of the probability of the electron to create a pair is different from that of photon essentially because, on one hand, the motion of the electron would be influenced by the electromagnetic field of the laser, on the other hand, the electron must first emit high energy photons randomly, and then, the emitted photons produce a pair probably. Accordingly, the probability of electron can not be obtained directly and should be obtained by a statistic method for a large number ($10^{5}$) of electrons. We also show the polynomial fitting of the statistic probability of the electron by the solid lines in Fig.\ref{fig.2.}.

As we have shown, the seed with higher incident energy has larger probability to create a pair by interacting with more powerful laser. However, the probability of photon is much larger than that of electron when the same initial condition is given. Because the field in the rest frame of electron would become much smaller than the field in the laboratory frame when the electron would be accelerated along the propagating direction of the laser (see Fig.\ref{fig.4.}). While the field in the center-of-mass system of Breit-Wheeler process is much larger than that in the laboratory frame as the photon would not be influenced by the electromagnetic field and would propagate countering to the laser.

\section{probability for emission of different particles}
\label{secIV}

In this section, we will discuss the probability ($W_{em}(t)\emph{d}t$) of the pair created by the seed-photon to emit photons. However, because of randomness, we can only take some samples to show the characteristics.

\begin{figure}[htbp]\suppressfloats
  \includegraphics[width=6cm]{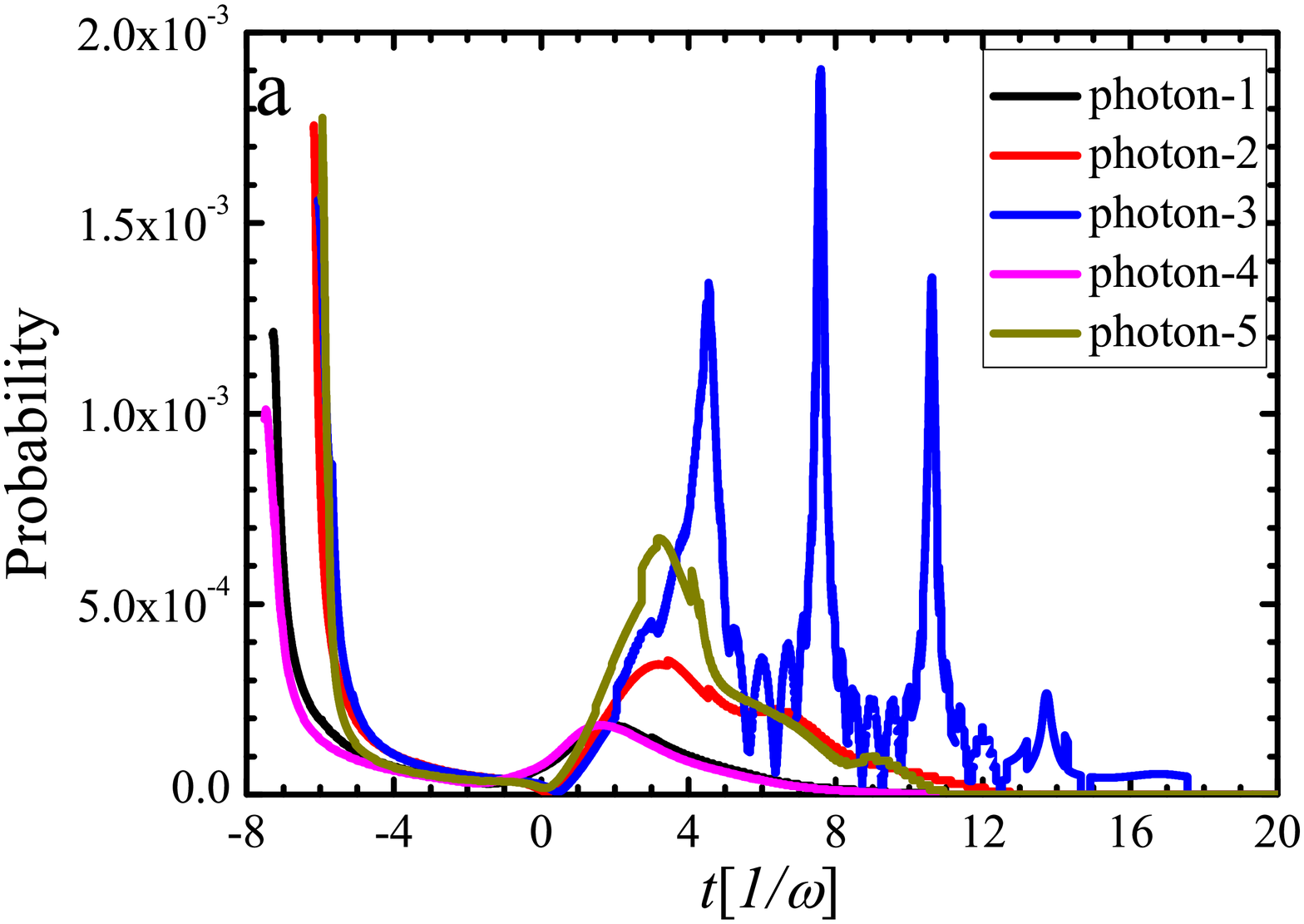}
  \includegraphics[width=6cm]{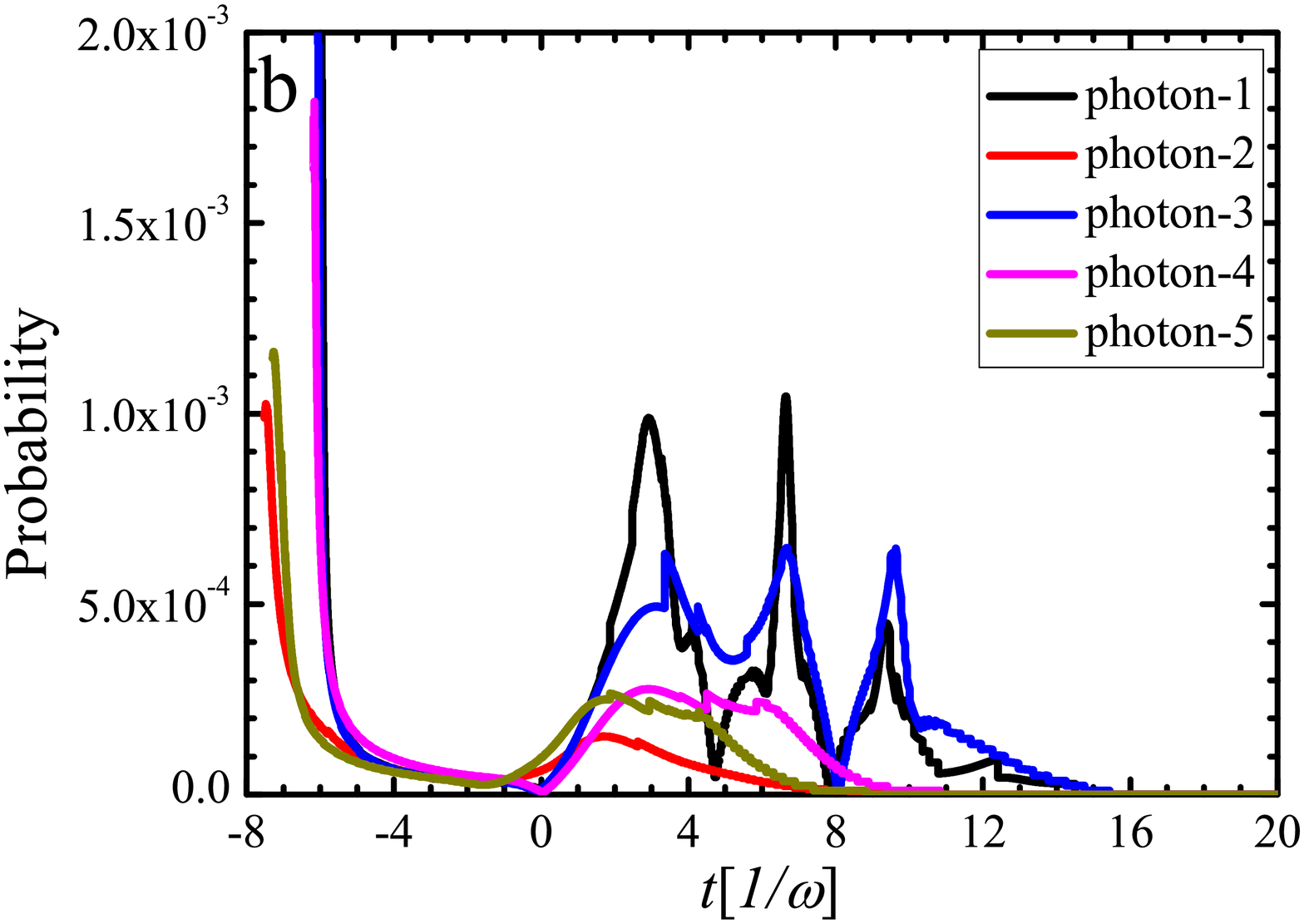}
  \includegraphics[width=6cm]{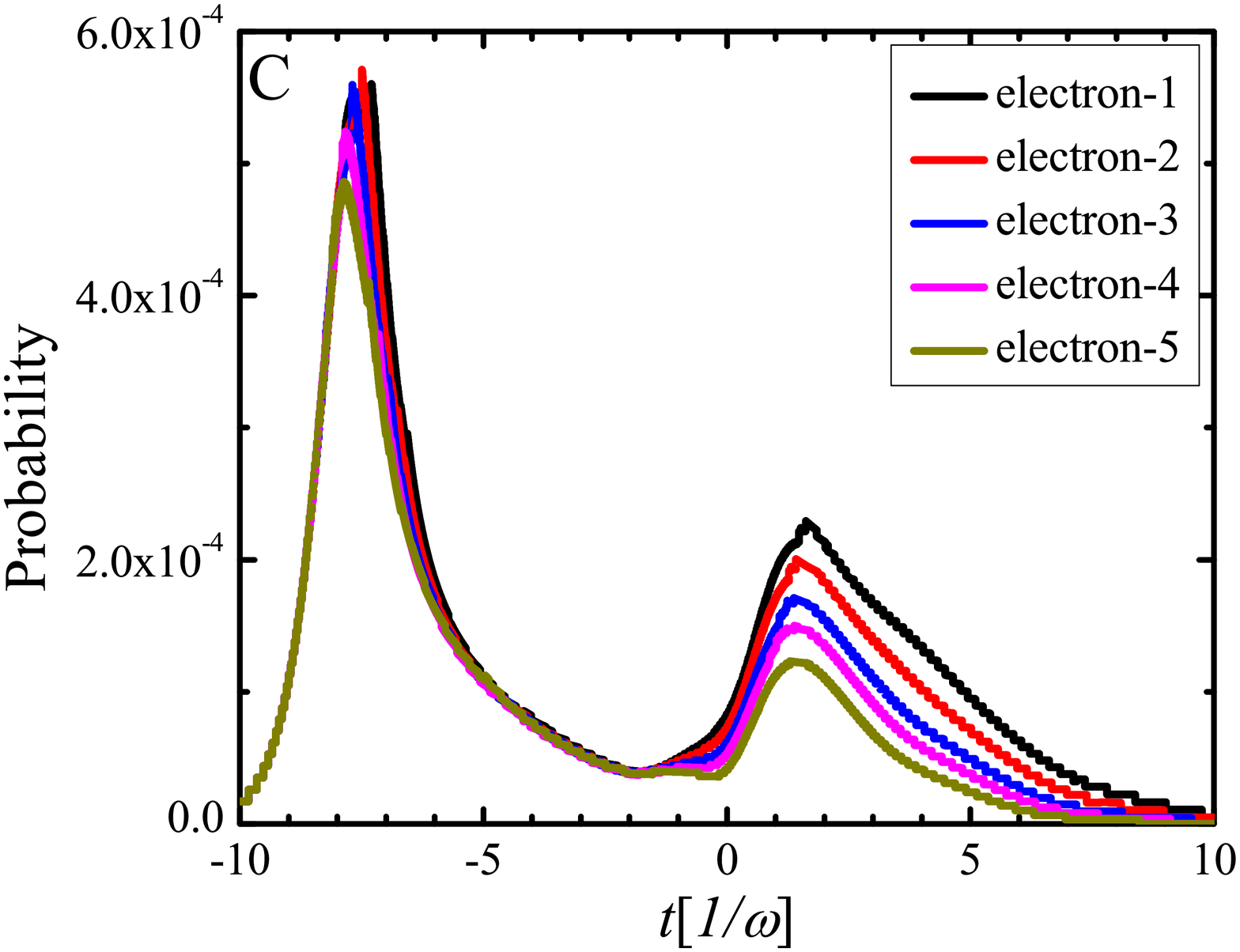}
  \caption{(Color online) The probability for emission of the electrons (a) and positrons (b) created by the seed-photons $\varepsilon=400$ and that of the seed-electrons (c) $\gamma=400$ with $I=5\times10^{24}\mathrm{W/cm^{2}}$. Because of randomness, the pairs are created at different time and the probability of the different sample is different.}\label{fig.3.}
\end{figure}

In Fig.\ref{fig.3.}, we show the probability of the electron (Fig.\ref{fig.3.}(a)) and positron (Fig.\ref{fig.3.}(b)) created by the seed-photon and that of the seed-electron (Fig.\ref{fig.3.}(c)) for emission with incident energy $\varepsilon,\gamma =400$ in the laser intensity $I=5\times10^{24}\mathrm{W/cm^{2}}$. As we can see, though the five pairs are created at different time by the photons randomly and the probability of different samples is very different from each other, the average probability of the created pair for emission is much larger than that of the seed-electron, i.e. the pair created by the seed-photon are more probable to emit photons than the seed-electron.

\begin{figure}[htbp]\suppressfloats
  \includegraphics[width=6cm]{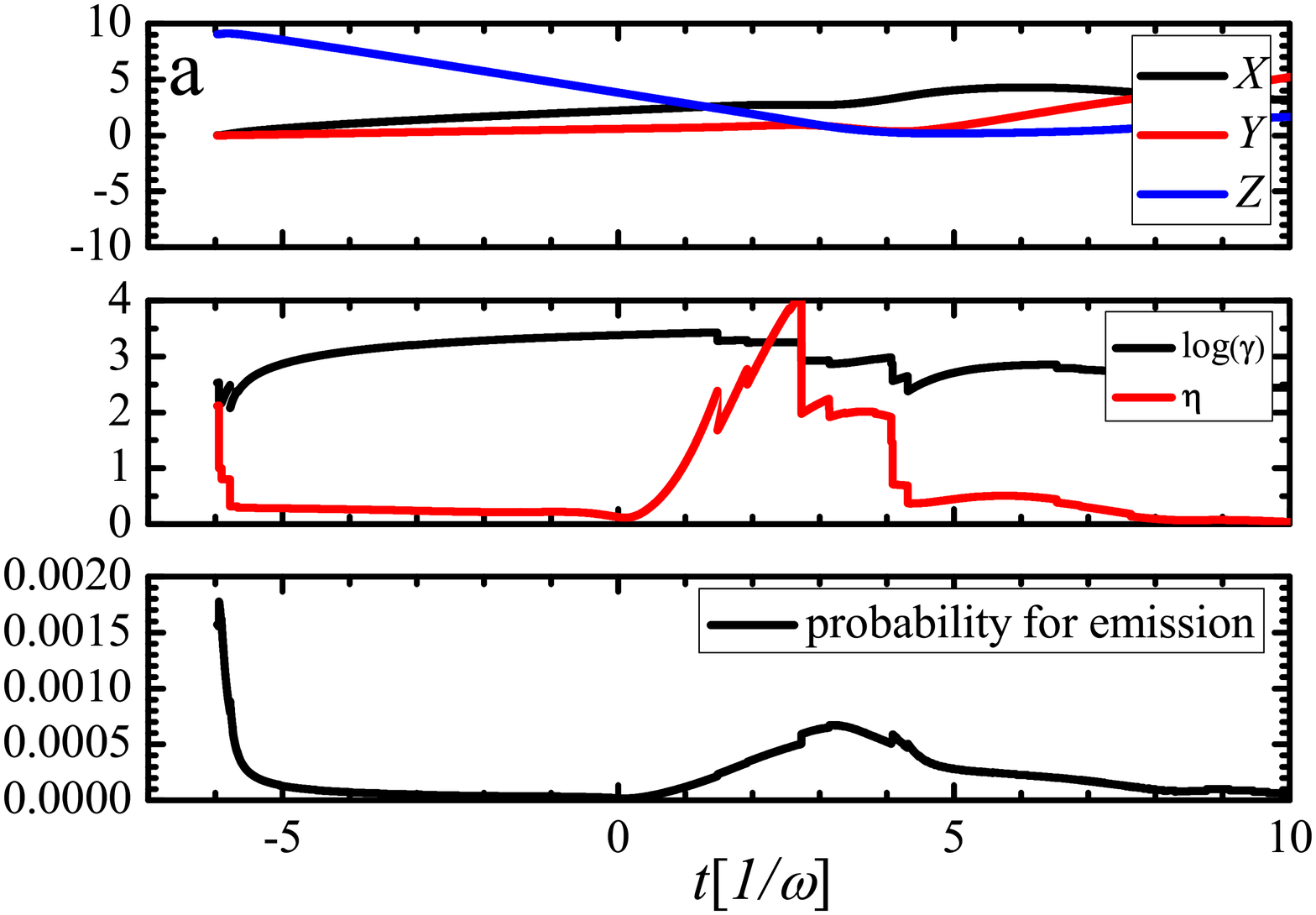}
  \includegraphics[width=6cm]{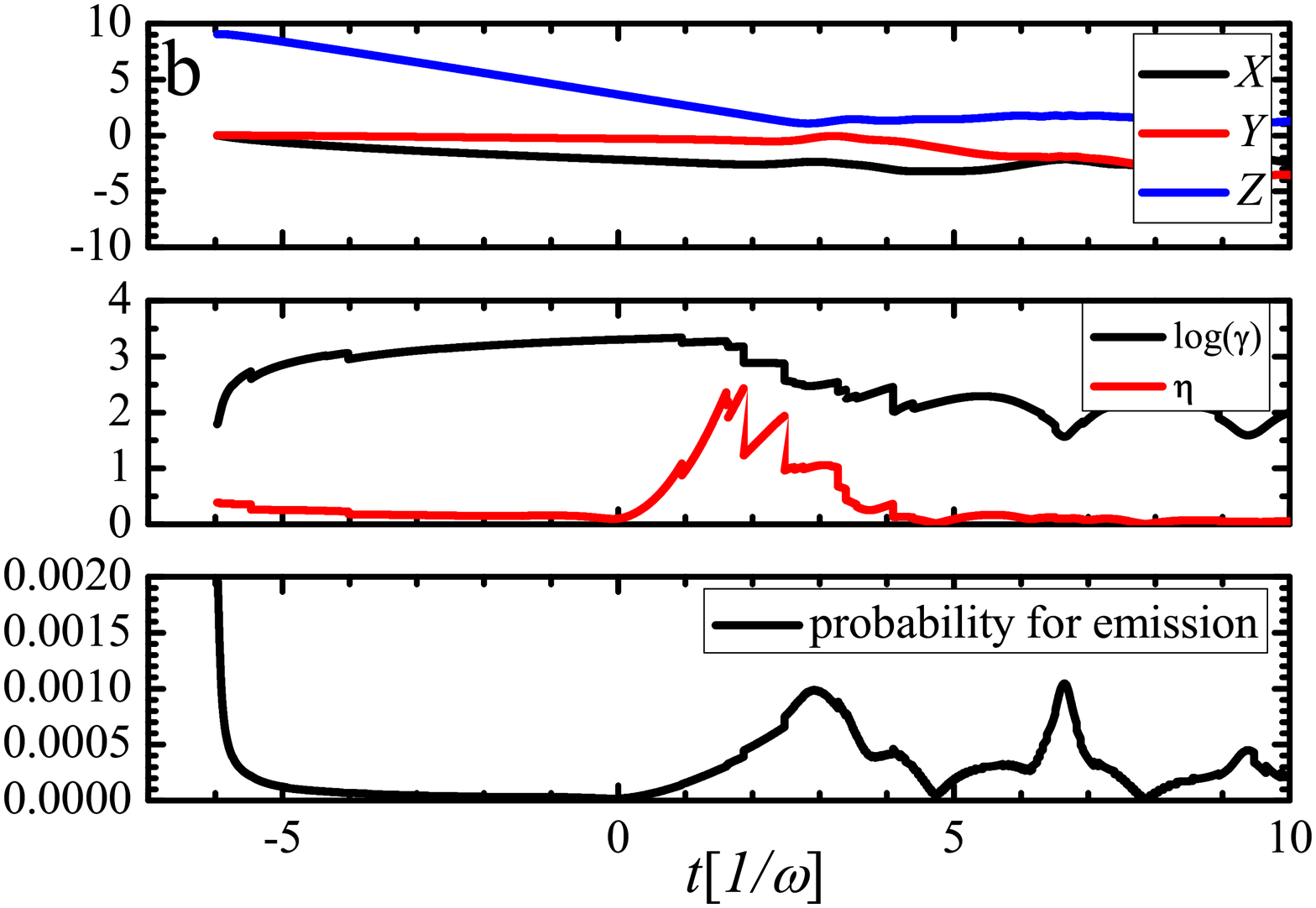}
  \includegraphics[width=6cm]{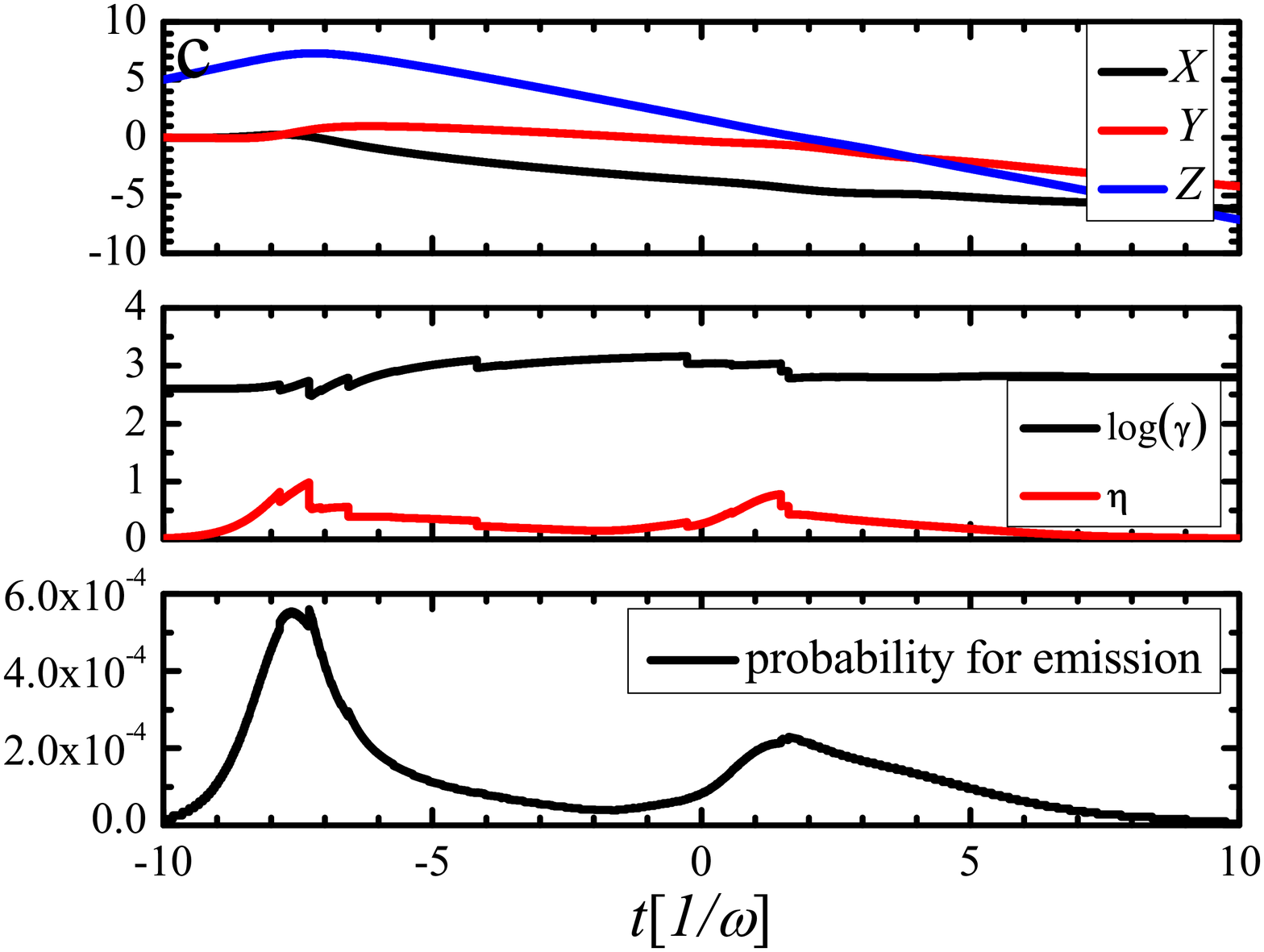}
  \caption{(Color online) The dynamic characteristics of the pair created by a seed-photon $\varepsilon=400$ (created electron (a), created positron (b))and that of a seed-electron $\gamma=400$ (c) with $I=5\times10^{24}\mathrm{W/cm^{2}}$. The pair is created at $z=9.026$ and $t=-5.97$ and the initial energy of the created electron is $\gamma_{e0}=338$ (a) and the energy of the positron is $\gamma_{p0}=62$ (b), thus, we can have energy conservation as $\gamma_{e0}+\gamma_{p0}=\varepsilon$.}\label{fig.4.}
\end{figure}

In order to demonstrate that the pair created by the seed-photon has general advantage in emitting photons compared to the seed-electron, we analyze the dynamic characteristics, which determine the probability for emission, of the created pair and compare with that of the seed-electron. In Fig.\ref{fig.4.}(a), (b), we illustrate the dynamic parameters of the created electron and positron. The electron and positron are created at time $t=-5.97$ and position $z=9.026$ in the leftward laser, and the energy is conserved as $\gamma_{e0}+\gamma_{p0}=\varepsilon$. After creation, the particles are accelerated to the left and encounter with rightward laser at time $t=1$. Because the electron and positron collide with the laser with large energy and the field of the laser at the colliding position (close to the axis) is strong, the parameter $\eta$ of the created electron and positron is much larger than $1$ during the encounter, thus the probability for emission is much larger than that of the seed-electron, see Fig.\ref{fig.4.}(c), which shows the dynamic parameters of the seed-electron in the interaction. The rightward electron encounters with the leftward laser at time $t=-7.5$. Then it is pushed back by the ponderomotive force and accelerated in the transverse direction by the electromagnetic field in the front of the laser. And at time $t=1.5$ the electron encounters with the rightward laser when the electron has been far away from the axis. In this process, the parameter $\eta$ of the electron is less than $1$ though it has two peaks during the interaction, so that the probability to emit photons is much smaller than that of the created pair. It is necessary to emphasize that though the dynamic parameters of the different samples may be different because of randomness, the characteristics discussed above can remain the same for different samples.

\section{QED cascade}
\label{secV}

In this section, the characteristics of the QED cascade induced by the different kinds of seeds are studied. Because of randomness, a large number of electrons and photons are used as seeds in this section to ensure to trigger the cascade.

\begin{figure}[htbp]\suppressfloats
  \includegraphics[width=8cm]{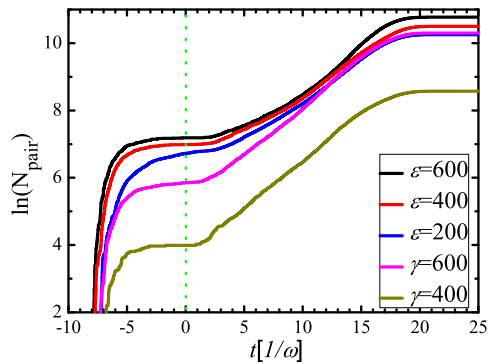}
  \caption{(Color online) The number of the created pairs $ N^{pair}$ in the cascade induced by the $1000$ seed-photons with incident energy $\varepsilon=600, 400, 200$ and the $1000$ seed-electrons with incident energy $\gamma=600, 400$ and the intensity of the laser is $I=5\times10^{24}\mathrm{W/cm^{2}}$. At time $t=0$, the number of the created pairs are $1320, 1080, 832$ for the seed photons and $347, 54$ for the seed-electrons}\label{fig.5.}
\end{figure}

In Fig.\ref{fig.5.}, we show the number of the pairs created in the cascade triggered by $1000$ seed-photons and electrons in the laser intensity $I=5\times10^{24}\mathrm{W/cm^{2}}$. The seeds move to the right from the initial position $z=-5$ with incident energy $\varepsilon=600,400,200$ and $\gamma=600,400$. As we can see, a number of pairs are created at time $t=-7.5$ when the seeds interact with the leftward laser, and it is obvious that the number of the pairs created by the seed-photons in this interaction is much larger than that by the seed-electrons as we have discussed above. Then, the created particles are accelerated to the left (see Fig.\ref{fig.4.}) and little new pairs can be created before the leftward particles encounter with the rightward laser (see Fig.\ref{fig.3.}). After the two lasers encounter at time $t=0$, the cascade is triggered and the number of the created pairs increases exponentially as $e^{\Gamma t}$ \cite{054401} until the two lasers depart from each other at time $t=20$, where $\Gamma$ is the increasing rate of the number of the created pairs.

\begin{figure}[htbp]\suppressfloats
  \includegraphics[width=8cm]{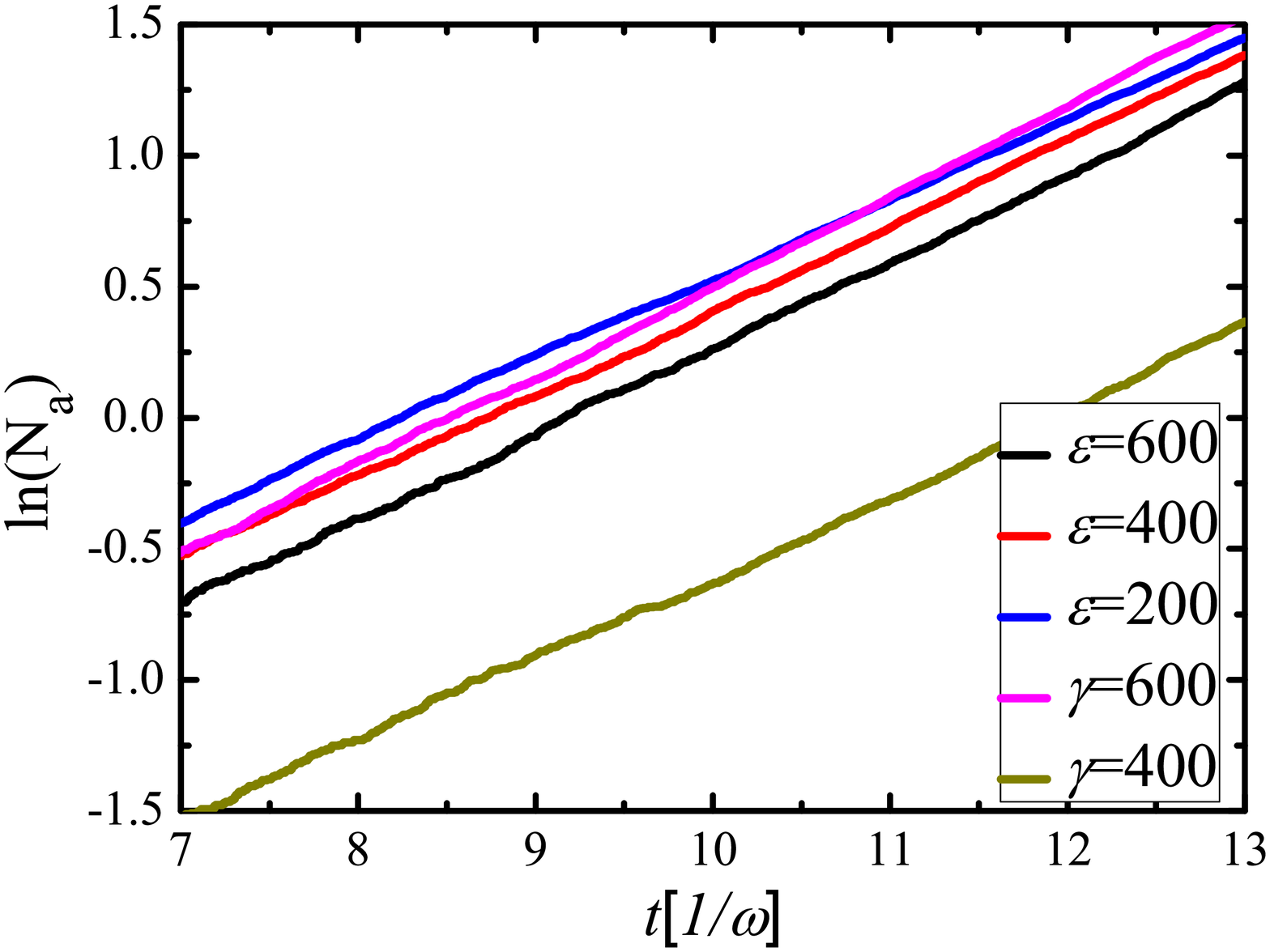}
  \caption{(Color online) The average number of the created pairs $N_{a}$ in the cascade induced by a charged particle with $I=5\times10^{24}\mathrm{W/cm^{2}}$. To ensure the linear relationship, we just match the values in Fig.\ref{fig.5.} in the interval between time $t=7$ and $t=13$. Before or after this interval, the linear property would be lost as the overlap region of the two lasers becomes smaller and smaller.}\label{fig.6.}
\end{figure}

In order to show the increasing rate $\Gamma$ of the created pairs clearly, we employ the approximate formula as $(N^{pair}(t)-N^{pair}_{0})/N_{0}=Ce^{\Gamma t}$ which can also be interpreted as the average number of the created pairs $N_{a}$ in the cascade induced by a charged particle, i.e., $N_{a}(t)=(N^{pair}(t)-N^{pair}_{0})/N_{0}$ and $\ln(N_{a}(t))=\ln(C)+\Gamma t$, where $N^{pair}_{0}$ and $N_{0}$ indicate the number of the created pairs and the charged particles at time $t=0$. We can have $N_{0}=2 N^{pair}_{0}$ if the seeds are photons and $N_{0}=2 N^{pair}_{0}+1000$ if the seed are electrons. From Fig.\ref{fig.5.}, we can get the values of $N^{pair}_{0}$ and therefore the values of $N_{0}$ can be obtained for each case as shown in Table \ref{table}. To exhibit the linear relationship between the average number $\ln(N_{a})$ and time $t$ clearly, we match the values in Fig.\ref{fig.5.} between the time interval $t=7$ to $t=13$ in which most of the particles are in the overlap region of the two lasers, and we can obtain the increasing rate $\Gamma$ as the slope shown in Fig.\ref{fig.6.}. Before or after the interval, the linear property would be lost as the overlap region of the two lasers is smaller which we do not shown.

\begin{table}
\caption{The values of $N^{pair}_{0}$ and $N_{0}$ from the Fig.\ref{fig.5.} and the linear fitting values of $\Gamma$ and $\ln(C)$ from Fig.\ref{fig.6.}.}\label{table}
\begin{ruledtabular}
\begin{tabular}{>{$}c<{$}>{$}c<{$}>{$}c<{$}>{$}c<{$}>{$}c<{$}>{$}c<{$}}

   & \varepsilon=600 & \varepsilon=400 & \varepsilon=200 & \gamma=600 & \gamma=400 \\\hline
  N^{pair}_{0} & 1320 & 1080 & 832 & 347 & 54 \\
  N_{0} & 2640 & 2160 & 1664 & 1694 & 1108 \\
  \Gamma & 0.3285 & 0.3199 & 0.3046 & 0.3423 & 0.3114 \\
  \ln(C) & -3.017 & -2.7860 & -2.5146 & -2.9207 & -3.7187 \\
\end{tabular}
\end{ruledtabular}
\end{table}

From the Table \ref{table}, we find that the increasing rates $\Gamma$ are almost the same for the different cases and obtain the average increasing rate as $\Gamma_{a}=0.3213$. And the difference between the intercept $\ln(C)$ of each line is negligible except the solid line representing $\gamma=400$. That is to say, the parameters $\Gamma$ and $\ln(C)$ do not depend on the kind of the seeds and are the function of the intensity of the laser \cite{054401}. The reason for the line of $\gamma=400$ far away from others can be understood easily as that most of the seed-electrons may not be influenced by the rightward laser because the seeds have been far away from the axis when they encounter with the rightward laser (see Fig.\ref{fig.4.}), i.e., the number of the effective charged particles at time $t=0$ is smaller than $N_{0}=1108$.

Since there is no difference in the values of $\Gamma$ and $C$ for each cascade induced by different seeds in the laser with the same intensity, the number of the pairs $N^{pair}$ created in the cascade can only depend on the number of the charged particles $N_{0}$. Consequently, we can conclude that the seed-photons are more probable to trigger QED cascade as much more pairs can be created by the seed-photons than by the seed-electrons before the two lasers encounter.

\section{summary}
\label{secVI}

In this paper, we study the QED cascade in the realistic situation in order to provide a reference for the future experiment. Two counter-propagating lasers are employed, and the photons and electrons are applied as the seeds to induce the cascade.

First we show that the probability of a seed-photon to create a pair is much larger than that of a seed-electron under the same condition. And second by analyzing the dynamic parameters of the pair created by the seed-photon, we find that the electron and positron created by the seed-photon have larger probability to emit photons than the seed-electron. Hence, we can deduce that more pairs can be created in the leftward laser induced by the seed-photon than by the seed-electron, i.e., there would be more charged particles stayed in the laser field before the two lasers overlap if the photon is used as the seed. Then we demonstrate that the number of the charged particles in the lasers at time $t=0$ when the two lasers overlap determines the number of the pairs created in the cascade because the increasing rate $\Gamma$ and constant $C$ only depend on the intensity of the laser. With all of these, we can have the conclusion that the QED cascade can be induced easier by the seed-photon than by the seed-electron.

We believe that our results presented here are not only helpful to deepen the understanding of the QED cascade with different seeds but also useful to open a possible way to obtain a good quality positron source which is important to many applications and would be report elsewhere.

\begin{acknowledgments}
This work was supported by the National Natural Science Foundation of China (NNSFC) under the grant No.11175023, and partially by the Fundamental Research Funds for the Central Universities (FRFCU). The numerical simulation was supported by the HSCC of Beijing Normal University.
\end{acknowledgments}

\end{document}